\documentclass[a4paper,11pt]{article}

\usepackage{amsmath,amssymb,graphicx}
\usepackage{latexsym}
\usepackage{epstopdf}
\usepackage{feynmf}		%Package for feynman diagrams.
\usepackage[font=small]{caption}

\def\ba{\begin{array}}
\def\ea{\end{array}}
\def\ba{\begin{eqnarray}}
\def\ea{\end{eqnarray}}

\def\lb{\label}
\def\be{\begin{equation}}
\def\ee{\end{equation}}
\def\theequation{\arabic{section}.\arabic{equation}}
\begin{document}

\baselineskip0.25in
\title{Probing global aspects of a geometry by the self-force on a charge:
Spherical thin-shell wormholes}
\author{E. Rub\'in de Celis$^{1}$\thanks{e-mail: erdec@df.uba.ar}\,, O. P. Santill\'an$^{2}$\thanks{e-mail: osantil@dm.uba.ar}\; and C. Simeone$^{1}$\thanks{e-mail: csimeone@df.uba.ar}\,.\\
\\ \\
{\small $^1$ Departamento de F\'isica, Facultad de Ciencias Exactas y Naturales,} \\
{\small Universidad de Buenos Aires and IFIBA, CONICET, Cuidad Universitaria, Buenos Aires 1428, Argentina.} \\
\\
{\small $^2$ CONICET--Instituto de Investigaciones Matem\'aticas Luis Santal\'o,}\\
{\small Ciudad Universitaria Pab. I, Buenos Aires 1428, Argentina.}}
\maketitle

\begin{abstract}
The self-interaction for a static point charge in the space-time of a thin-shell wormhole constructed connecting two identical Schwarzschild geometries is calculated in a series expansion. The electrostatic self-force is evaluated numerically. It is found to be attractive towards the throat except for some values of the throat radius proximate to the value of the Schwarzschild horizon for which the force is repulsive or attractive depending on the position of the charge. The result differs from the self-force in the space-time of the Schwarzschild black hole, where it is always repulsive from the center. Although these wormhole and black hole geometries are locally indistinguishable, the different topologies of both backgrounds are manifested in the electrostatic field of a point charge.\\

\end{abstract}

\section{Introduction}

Electrodynamics in General Relativity is described by the Maxwell equations in curved space-time \cite{ll}. A freely falling observer in such background would write the same equations valid for Minkowski space-time; however, these equations must have a different solution, because the curved geometry imposes a different asymptotic behavior than the flat one. In particular, the electric field around a static point charge in a curved background is not spherically symmetric in general, and this has the consequence of a so-called electrostatic self-force on the charge.

One of the earliest studies on the electrostatic self-force on static charges induced by a curved background was that on a black hole geometry. It was shown that the self-force on a charge $q$ is repulsive, i.e. it points outwards from the black hole, and that it has the dependence
\be
f\sim\frac{mq^2}{r^3},
\ee
where $m$ is the mass of the source and $r$ is the Schwarzschild radial coordinate of the charge. This result was first obtained within the framework of linearized general relativity \cite{vil79}, and was later recovered working within the full theory \cite{will80}. After the publication of these leading works the study of the self-interaction of a charge was extended to other geometries.  A notable result was the self-force on a charge in the vicinity of a straight cosmic string arising from symmetry breaking in a system composed by a complex scalar field coupled to a gauge field \cite{lin}. The associated geometry is locally flat but includes a deficit angle determined by $\mu$, the mass per unit length of the string \cite{vilgothis}. The self-force in this case points outwards from the cosmic string and is proportional to $\mu/r^2$. This non null self-force in a locally flat background is of great interest because it shows how the global properties of a manifold (in this case, the existence of a deficit angle) are revealed by the electromagnetic field of the charge. In fact, these results together with the calculation of the self-force on a point charge in a wormhole space-time \cite{kb}, which turned out to be attractive, i.e. towards the wormhole throat, suggested the possibility of detecting thin-shell wormholes by means of electrostatics. Differing from well-known wormholes of the Morris--Thorne type \cite{mo} which are supported by non localized exotic matter, thin-shell wormhole geometries are supported by a shell of exotic matter located at the wormhole throat \cite{visser}. The throat connects two (equal or different) geometries which can be those of other astrophysical objects. For example, Schwarzschild thin-shell wormholes connect two  exterior (that is, beyond the horizon) non charged black hole space-times; hence the geometry at each side of the throat is locally identical to the exterior of a black hole geometry. However, the topology of the wormhole geometry is non trivial, thus the  global properties are essentially different in each case. Our proposal is that global aspects, such as the existence of a throat or not, can be revealed by electrodynamics, more precisely, by the electrostatic self-force on a point charge. In our recent article \cite{eoc12} we developed this proposal and applied it to the case of wormholes with a cylindrical throat which are mathematically constructed by removing the regions $r<a$ of two gauge cosmic string manifolds and pasting the two regions $r\geq a$. We obtained the self-force on a charge in the cylindrical wormhole geometry, and compared it with the self-force on a charge in the vicinity of a gauge cosmic string. We showed that the force in the wormhole case can be attractive or repulsive depending of the position of the charge; this result would then allow an observer to distinguish between two geometries which are locally equal.

It is interesting to remark that there exist some works related to these ideas. For instance, in \cite{gralla}, the authors considered a minimally coupled scalar charge and a electromagnetic charge when a Schwarzschild black hole interior is replaced by a material body and found that the leading term in a large-$r$ expansion of the force was independent of the central body type. Nevertheless, when the scalar charge is not minimally coupled, the self-force is dependent on the composition of the body. Another work in the same line is \cite{poisson}, where a spherical ball of perfect fluid in hydrostatic equilibrium with rest mass density and pressure related by some polytropic equations of state is considered. The authors found that the leading term of the force is universal and does not distinguish the internal body structure, but the next-to-leading order term is sensible to the equation of state. Thus the self-force distinguishes the body composition. In the present work we extend our study by applying our proposal to the physically interesting case of a Schwarzschild thin-shell wormhole: we consider a static point charge in the topologically non trivial space-time constituted by two Schwarzschild geometries connected by a wormhole throat (with the throat radius larger than the Schwarzschild horizon radius), and we compare the result with the self-force on a charge in the vicinity of a non charged black hole. While some aspects of the analysis will be very similar to those in \cite{eoc12}, we will see that the different asymptotic behaviors presented by the cosmic string and black hole geometries will be reflected in some interesting differences in the results.\\

\section{Field of an electrostatic charge in a thin-shell Schwarzschild wormhole}

\setcounter{equation}{0}  % reset counter

In the following the electrostatic potential of a static charge in front of a Schwarzschild black hole will be considered as a series expansion. The same expansion will be used to calculate the potential of the charge in presence of a Schwarzschild thin-shell wormhole. The metric for a black hole space-time in Schwarzschild coordinates is given by the line element
\be\lb{S}
ds^2= -\left(1-\frac{2m}{r}\right)dt^2+\left(1-\frac{2m}{r}\right)^{-1}\,dr^2+r^2\left(d\theta^2+\sin^2\theta d\varphi^2\right)\,,
\ee
with $t$ taking values in $(-\infty, +\infty)$, $\theta$ in [$0$, $\pi$), $\varphi$ in [$0$, $2\pi$) and $r$ is the radial coordinate, which takes values in $\mathbb{R}_{>0}$.  The horizon of the black hole is located at $r=2m$. In order to fix notations and conventions, we recall that the Maxwell equations are given by
\be\lb{max}
4\pi j^{\alpha}=F^{\mu \alpha}_{\quad ;\mu}\;,
\ee
with the electromagnetic tensor given by $F_{\mu\nu}=\partial_{\mu}A_{\nu}-\partial_{\nu}A_{\mu}$. For a point charge $q$ held at rest at $r=a>2m$ and $\theta=0$ in the black hole space-time, the covariant time-like component of the electromagnetic four-potential, $A_t$, will represent an electrostatic potential which we will call $A_t=V^{bh}$. The Maxwell equations in this case reduce to
\be\lb{eq}
-4\pi \rho=\frac{1}{r^2}\partial_r\left(r^2\partial_rV^{bh}\right)+\frac{1}{r^2\sin\theta}\frac{1}{\left(1-2m/r\right)}
\partial_{\theta}\left(\sin\theta\partial_{\theta}V^{bh}\right),
\ee
where
\be\lb{rho}
\rho=q\frac{\delta(r-a) \delta(\theta)}{2\pi r^2 \sin\theta},
\ee
 is the time-like component of the four-current $j^{\alpha}$. The other components are $A_i=0$ for $i=r,\theta,\varphi$. With the previous definitions the electrostatic field is calculated as
\be\lb{field}
(\vec{E})^i\equiv-F^{ti}=-g^{tt}g^{ii}F_{ti}=g^{tt}g^{ii}\partial_i A_t \,.
\ee
Since we are working in spherical coordinates, the potential can be expanded as
\be\lb{Vbh}
V^{bh}=\sum_{l=0}^\infty R_{l}(r)P_{l}(\cos \theta)\,,
\ee
where $P_{l}(\cos \theta)$ are the standard Legendre polynomials and the radial functions $R_l(r)$
are solutions of the homogeneous equation
\be\lb{lapli}
\left(1-\frac{2m}{r}\right)\frac{d}{dr}\left(r^2 \frac{dR_l}{dr}\right)-l(l+1)R_l=0\,.
\ee
The two independent solutions of (\ref{lapli}) are \cite{israel}
\be\lb{fyg}
\begin{array}{ll}
f_{l}(r)=-\frac{(2l+1)!}{2^l (l+1)!l!m^{l+1}}\,(r-2m)\,\frac{dQ_l}{dr}\left(\frac{r}{m}-1\right)\,.\\\\
g_{l}(r)=\left\{
\begin{array}{ll}
  1 \,,\qquad \textrm{for} \;\, l=0\,. \\
  \frac{2^l l! (l-1)! m^l}{(2l)!}\,(r-2m)\,\frac{dP_l}{dr}\left(\frac{r}{m}-1\right)\,,\qquad \textrm{for} \; \, l\neq 0 \,.
\end{array} \right.
\end{array}
\ee

Here $P_l(x)$ and $Q_{l}(x)$ are the two types of Legendre functions. These solutions possess the following asymptotic behavior when $r\to \infty$:
\be\lb{asym}
g_l(r)\to r^l,\qquad f_l(r)\to 1/r^{l+1}.
\ee
These limits corresponds to the solutions of a standard problem of electrostatics in flat backgrounds \cite{jackson}, which is plausible since the effect of the black hole is washed out at large distances. On the other hand, as $r\to 2m$; $f_l(r) \to$ finite constant, $f_l'(r) \sim \log(1-2m/r)$ for $l\neq 0$, and $g_l(r) \to 0$ while $g'_l(r)\to$ constant. Note that $f_0=r^{-1}$ and $g_0=1$.

The electrostatic potential in series expansion corresponding to the black hole geometry was calculated by Cohen and Wald in \cite{cohen-wald} and is given by \footnote{We have defined Maxwell equations with the sign given in (\ref{max}) to obtain the same expansion, (\ref{Vcw}), as in \cite{cohen-wald} with the functions (\ref{fyg}).}

\be\lb{Vcw}
V^{bh}=\left\{
\begin{array}{ll}
q \sum_{l=0}^\infty g_{l}(a)f_{l}(r)P_{l}(\cos\theta) \, , \qquad \textrm{for} \;\, r \geq a\,. \\\\
q \sum_{l=0}^\infty g_{l}(r)f_{l}(a)P_{l}(\cos\theta) \, , \qquad \textrm{for} \;\, r \leq a \,.
\end{array} \right.
\ee
\\
Let us turn now our attention in computing the electrostatic potential of the same charge $q$ in the space-time of a thin-shell Schwarzschild wormhole. This wormhole is constructed taking two copies of the Schwarzschild geometry and removing from them the four-dimensional regions described by
\be
\Omega_{1,2}=\{r_{1,2}\leq c \mid c>2m\}\, .
\ee
One is left with two identical geodesically incomplete manifolds with boundaries given by the time-like hypersurfaces:
\be
\partial \Omega_{1,2} = \{ r_{1,2}=c \mid c>2m\}\, .
\ee
Identifying these two hypersurfaces (i.e., $\partial \Omega_1=\partial \Omega_2$) of each copy, the resulting space-time is complete and possesses two asymptotically flat regions connected by a wormhole \cite{visser}. Note that the condition $c>2m$ is necessary to prevent the formation of the event horizon. At the throat of the wormhole, $\partial \Omega$, the stress-energy tensor is proportional to a delta function representing a thin layer of exotic matter. This thin-shell wormhole mathematical construction is based on the junction condition formalism, which is one of the major tools for studying traversable wormholes. The metric for this thin-shell Schwarzschild wormhole is given by the line element:
\be
ds^2 = -\left(1-\frac{2m}{r_{1,2}}\right)dt^2+\left(1-\frac{2m}{r_{1,2}}\right)^{-1}\,dr_{1,2}^2+r_{1,2}^2\left(d\theta^2+\sin^2\theta d\varphi^2\right)\, ,
\ee
with $r_1,r_2 \geq c >2m$ and the other coordinates defined as before. The potential for a static charge $q$ located at $r_1=a$ and $\theta=0$ in this space-time has a spherical expansion as in (\ref{Vbh}) with azimuthal symmetry. The general form which is not divergent at any of both infinite regions is given by
\be\lb{Vwh}
V^{wh}_{1,2}=\left\{
\begin{array}{ll}
V^{wh}_1=\left\{
\begin{array}{ll}
\sum_{l=0}^{\infty} A_l f_{l}(r_1)P_{l}(\cos\theta)\,,\qquad \qquad \qquad \, \, \textrm{for} \;\, r_1 \geq a\,. \\\\
\sum_{l=0}^{\infty} \left(C_l f_{l}(r_1)+D_l g_l(r_1)\right)P_{l}(\cos\theta)\,,\qquad \textrm{for} \; \, c \leq r_1 \leq a \,.
\end{array} \right.\\\\
V^{wh}_2=\sum_{l=0}^{\infty} E_l f_{l}(r_2)P_{l}(\cos\theta)\,,\qquad \qquad \qquad \, \, \textrm{for} \;\, r_2 \geq c\,.\\
\end{array}\right.
\ee

The subindex $1$ refers to the region of the wormhole where the charge is located at $r_1=a$ and the subindex $2$ to the complementary copy (the other, empty, Schwarzschild geometry). In each region the radial coordinate, $r_1$ and $r_2$ respectively, extend from $[c, +\infty]$ and if there is no confusion they will be referred to as $r$.
The potential is defined, up to an irrelevant constant, by the following boundary conditions:
\be\lb{cc1}
V^{wh}_1(r_1 \to a^-,\theta)=V^{wh}_1(r_1 \to a^+,\theta)=V^{wh}_1(a,\theta)\,,
\ee
\be\lb{cc2}
\frac{dV^{wh}_1}{dr}(r_1 \to a^-,\theta)-\frac{dV^{wh}_1}{dr}(r_1 \to a^+,\theta)=4\pi \frac{q\,\delta(\theta)}{2\pi a^2 \sin \theta}\,,
\ee
\be\lb{cc3}
V^{wh}_1(r_1 \to c^+,\theta)=V^{wh}_2(r_2 \to c^+,\theta, \varphi)=V^{wh}_{1,2}(c,\theta)\,,
\ee
\be\lb{cc4}
\frac{dV^{wh}_1}{dr_1}(r_1 \to c^+,\theta)=-\frac{dV^{wh}_2}{dr_2}(r_2 \to c^+,\theta)\, .
\ee
This conditions are simply the continuity of the potential at the radial location of the charge (\ref{cc1}) and at the throat (\ref{cc3}), together with the requirement that the discontinuity of the electric field when crossing the charge's radial location be proportional to $\frac{q\,\delta(\theta)}{2\pi a^2 \sin^2 \theta}$, the surface charge density at $r_1=a$ due to the point charge, (\ref{cc2}). Additionally, in (\ref{cc4}), the field is required to be continuous at the throat.

The first boundary condition (\ref{cc1}) implies that
\be\lb{3}
A_l f_{l}(a)=C_l f_{l}(a)+D_l g_l(a)
\ee
In addition, multiplying the second, (\ref{cc2}), by $P_l(\cos\theta)$ and integrating using the orthogonality relation
\be
\int_0^1 P_{l}(x)P_{k}(x)dx=\frac{2}{2l+1}\delta_{kl},
\ee
it is obtained that
\be
q\frac{(2l+1)}{a^2}=C_l f'_{l}(a)+D_l g'_l(a)-A_l f'_{l}(a).
\ee
By taking into account (\ref{3}) the last expression may be transformed to
\be
q\frac{(2l+1)}{a^2}f_{l}(a)=D_l (f_{l}(a)g'_l(a)- g_l(a)f'_{l}(a)).
\ee
This quantity can be evaluated by noticing that the right-hand-side is proportional to the Wronskian between $g_l$ and $f_l$, whose value is known to be \cite{cohen-wald}
$$
W(g_l(a); f_l(a); a)=g_l(a)f'_{l}(a)-f_{l}(a)g'_l(a)=-\frac{(2l+1)}{a^2}.
$$
Therefore,
\be\lb{4}
D_l=q\,f_{l}(a).
\ee
The last equality, combined with (\ref{3}), implies that
\be\lb{5}
A_l=C_l+q\,g_l(a).
\ee
In addition, the boundary conditions (\ref{cc3}) and (\ref{cc4}) constitute a simple linear algebraic system for $C_l$ and $E_l$,
the solution is given by
 \be\lb{6}
C_l =-q\frac{f_{l}(a)}{f_{l}(c)f'_{l}(c)}\bigg[\frac{(2l+1)}{2c^2}+f'_{l}(c)g_{l}(c)\bigg],
\ee
\be\lb{7}
E_l =-q\frac{f_{l}(a)(2l+1)}{2c^2f_{l}(c)f'_{l}(c)},
\ee
where the Wronskian value $W(g_l(c); f_l(c); c)=-(2l+1)/c^2$ was taken into account to derive these expressions. The last result, combined with
(\ref{5}) shows that
\be\lb{8}
A_l=-q\frac{f_{l}(a)}{f_{l}(c)f'_{l}(c)}\bigg[\frac{(2l+1)}{2c^2}+f'_{l}(c)g_{l}(c)\bigg]+q\,g_l(a).
\ee
Finally the electrostatic potential (\ref{Vwh}) for the probe charge
in the wormhole geometry is explicitly given by
\be\lb{cw2}
V^{wh}_{1,2}=\left\{
\begin{array}{ll}
V^{wh}_1=\left\{
\begin{array} {ll}
 q \sum_{l=0}^{\infty}  \left[g_l(a)-\left(\frac{(2l+1)}{2c^2f'_{l}(c)}+ g_{l}(c)\right)\frac{f_{l}(a)}{f_{l}(c)} \right] f_{l}(r) P_{l}(\cos\theta)\,, \qquad r_1 \geq a\,. \\\\
 q \sum_{l=0}^{\infty}  \left[f_l(a) g_l(r)-\left(\frac{(2l+1)}{2c^2f'_{l}(c)}+ g_{l}(c)\right)\frac{f_{l}(a)}{f_{l}(c)} f_{l}(r)\right]P_{l}(\cos\theta)\, ,\qquad c \leq r_1 \leq a\,.
\end{array}\right.\\\\
V^{wh}_{2}=-q \sum_{l=0}^{\infty} \frac{(2l+1)}{2c^2f'_{l}(c)} \frac{f_{l}(a)}{f_{l}(c)} f_{l}(r)P_{l}(\cos\theta)\, ,\qquad r_2 \geq c\,.
\end{array}\right.
\ee
\\
It is interesting to check the Gauss law for this solution. To this end one should consider observation points $r$ sufficiently far from the charge's position and the wormhole throat, that is $r \to \infty$ in both sides of the space-time of the wormhole. The Gauss theorem reads
\be\lb{Gauss}
4\pi q=\int\int_{S_1}\vec{E}_1\hat{n}_1dS_1+\int\int_{S_2}\vec{E}_2\hat{n}_2dS_2,
\ee
where $\vec{E}_j$ is the electrostatic field and $S_j$ is a surface enclosing the throat with exterior $3-dimensional$ normal vector $\hat{n}_j$ pointing towards the asymptotic infinity corresponding to the coordinate $r_j$. Therefore, when taking the limit $r_j \to \infty$ (for $j=1,2$), by virtue of the asymptotic behavior of the functions $f_l(r)$ mentioned in (\ref{asym}), only the first term $l=0$ of the electric field series survives. This represents the monopole term of the electric field in each region:
\be\lb{asymE}
\begin{array} {ll}
(\vec{E}_1)^r=-\partial_rV^{wh}_1 \sim q \left(1-\frac{c}{2a} \right) \frac{1}{r^2} \\\\
(\vec{E}_2)^r=-\partial_rV^{wh}_2 \sim q \left(\frac{c}{2a}\right) \frac{1}{r^2}
\end{array}
\ee
When these fields are introduced in (\ref{Gauss}) the Gauss theorem is exactly satisfied. The expressions (\ref{asymE}) represent the electric field flux density flowing to both infinities of the space-time as a function of the throat radius, $c$, and the position of the charge, $a$. Note that the flux is equally distributed when the charge is placed at the throat. This gives us confidence that the solution found is indeed the correct one, as it satisfies the desired boundary condition for the electric field at infinity.

We turn now our attention to the problem of calculating the self-force experienced by the static charge due to its own electrostatic field.\\

\section{Calculation of the electrostatic self-force in the wormhole geometry}
\setcounter{equation}{0}  % reset counter

The expression (\ref{cw2}) found above is clearly the sum of the terms in (\ref{Vcw}) plus contributions that arise due to the wormhole topology. The Schwarzschild part (\ref{Vcw}) can be summed to give (see appendix)
\be\lb{willy}
V^{bh}=V^{C}+V^{L}=\frac{q}{ar}\frac{(r-m)(a-m)-m^2\cos\theta}
{[(r-m)^2+(a-m)^2-2(r-m)(a-m)\cos\theta - m^2\sin^2\theta]^{1/2}}+\frac{q m}{ar}.
\ee
The first term, $V^{C}$, was derived by Copson in \cite{copson} following Hadamard's theory of elementary solutions for partial differential equations \cite{hadamard}. Copson's potential is a local construction and provides an exact solution to the field equations for the electrostatic potential sourced by the particle in Schwarzschild geometry which is singular at the position of the charge. It is also divergent in the limit $r \to 0$, a pathological consequence of the metric which wouldn't be a problem if the correct boundary conditions were satisfied. The second term, $V^{L}$, was added by Linet (\cite{Linet}) and is a homogeneous solution which insures that the boundary conditions at infinity for the electric field derived from $V^{bh}$ are fulfilled. In these terms, assuming the convergence of the series in (\ref{cw2}), the electrostatic potential in the region of the wormhole where the charge is located is given by
\be\lb{ba2}
V_1^{wh}= V^{bh} - q \sum_{l=0}^{\infty}  \left[\frac{(2l+1)}{2c^2f'_{l}(c)}+ g_{l}(c) \right] \frac{f_{l}(a)}{f_{l}(c)}f_{l}(r) P_{l}(\cos\theta).
\ee
When this expression is evaluated at the position of the charge, the potential diverges. The singular part is known to depend only on the local properties of the geometry in a neighborhood of the charge's position. In other words, we can say that the singular part of the wormhole's potential is the same as the one in the black hole case. Removing this singular potential at the position of the charge reveals a regular homogeneous solution from which the self-interaction can be computed. Several renormalization procedures have been used in the past in order to remove singular potentials. The renormalization method that enjoys the best justification is that of Detweiler and Whiting which is based in a four dimensional singular Green function \cite{DW}. Recently in \cite{Casals}, the authors showed the equivalence between this procedure in the case of a static particle in a static space-time and Hadamard's two-point function in three
dimensions for the computation of electrostatic self-forces  \footnote{The equivalence is guaranteed up to certain order in the Green's function expansion and conjectured to all order. Only the first terms are needed to compute the self-force, as shown in (\ref{Gs}), so they are said to be equivalent for this type of calculations.}. An alternative approach to renormalization, which is also suitable for charged particles at rest in general static curved space-times is the DeWitt-Schwinger asymptotic expansion of the three-dimensional Green function, which was considered recently in \cite{kpl}. In either of these formalisms, the renormalized potential at the position of the charge is given by
\be\lb{ren}
V^{wh}_{ren}(x^{i'})=\lim_{x^{i}\to x^{i'}}(V^{wh}-V^{sing})\,.
\ee
In the last equation the coincidence limit takes the coordinate spatial components $x^i$ to the charge's position $x^{i'}$ along the shortest geodesic connecting them. The singular term in this definition is
\be
V^{sing}=\sqrt{-g_{t't'}}G^s_3(x^i;x^{i'}),
\ee
where $G^s_3(x^i,x^{i'})$ is the singular Green function in three dimensions, and the primed indices refer to the position of the source charge. The Green function must have the same singularity structure as the particle's actual field and exert no force on the particle. The three methods mentioned above agree in the following expansion for the singular Green function (\cite{hadamard}, \cite{DW}-\cite{kpl}):
\be\lb{Gs}
G^{s}_3(x^i;x^{i'})=\frac{q}{\sqrt{2\sigma}}(1-\frac{g_{t't',i'}\sigma^{,i'}}{4g_{t't'}}
+\mathcal{O}\left(\sigma\right)),
\ee
where $\sigma=\sigma(x^i,x^{i'})$ is half the squared geodesic distance between $x^i$ and $x^{i'}$ as measured in the purely spatial sections of the space-time and $\sigma^{,i'}=g^{i'j'}\partial \sigma / \partial x^{j'}$ (see Refs. \cite{Casals} or \cite{kpl}  for a full derivation\footnote{In (\ref{Gs}) there is an overall sign difference with respect to usual literature arising from the convention taken for the Maxwell equations, (\ref{max}).}). In the expansion (\ref{Gs}) the terms of order $\mathcal{O}\left(\sigma\right)/\sqrt{2\sigma}$ are irrelevant for the renormalization of the potential field since they vanish in the coincidence limit taken in (\ref{ren}).

To calculate (\ref{ren}) we can evaluate at coincidence angles in advance, i.e. $\theta=0$, and take the limit as $r$ approaches $a$ from the right using (\ref{willy}) to express $V_1^{wh}$ for $r \geq a$:
\be\lb{v}
V^{wh}=V_1^{wh}(r \geq a,\theta=0)=\frac{q}{|r-a|}\left(1-\frac{2m}{r}\right)- q \sum_{l=0}^{\infty}  \left[\frac{(2l+1)}{2c^2f'_{l}(c)}+ g_{l}(c) \right] \frac{f_{l}(a)}{f_{l}(c)}f_{l}(r)\, .
\ee
When $x^{r'}=a$ and $x^{\theta'}=0$, $\sigma$ is one half of the squared radial geodesic distance between $r$ and $a$,
\be\lb{sigma}
\sigma(r,a)=\frac{1}{2}\left(\int^a_{r}\frac{dr'}{\sqrt{1-\frac{2m}{r'}}}\right)^2,
\ee
which is completely defined in terms of the local properties of the Schwarzschild metric.
Using (\ref{v}) and (\ref{sigma}) to compute explicitly (\ref{ren}) by taking the limit $r \to a^+$, the renormalized potential at the position of the charge is obtained;
\be\lb{Vren}
V^{wh}_{ren}(a)=\frac{q m}{a^2}-q \sum_{l=0}^{\infty}  \left[\frac{(2l+1)}{2c^2f'_{l}(c)}+ g_{l}(c) \right] \frac{f_{l}(a)^2}{f_{l}(c)},
\ee
Observe that this result can be directly picked up from (\ref{ba2}) removing Copson's solution, $V^{C}$, from $V^{bh}$ and evaluating in the position of the charge. What we had checked out in this calculation is that
\be
\lim_{x^{i}\to x^{i'}}(V^{C}-V^{sing})=0\,,
\ee
so the general singular Green function for the point charged particle at rest in this static geometry removes the singular part of the potential and exerts no force as expected. In other words, Hadamard's elementary solution constructed by Copson coincides with the Green function.

\begin{figure}[!t]
\begin{center}
  \includegraphics[width=11cm]{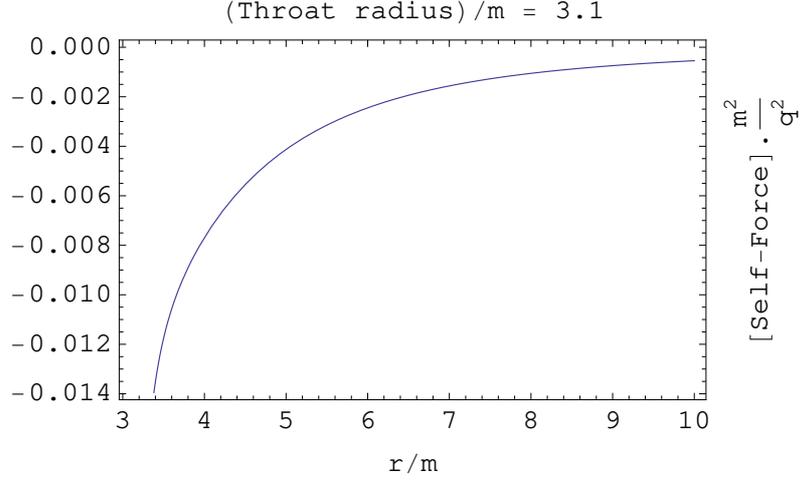}
  \caption{Dimensionless radial self-force ($\left(\mathrm{f}^{self}\right)\frac{m^2}{q^2}$) as a function of the dimensionless coordinate $r/m$ of a charge $q$ when the throat radius is at $c=3.1m$. The graph represent values for the position $r$ in the range $(c=3.1m,10m)$. Only attractive electrostatic self-force is possible in this case despite its throat is very close to 2m.}\label{3.1}
\end{center}
\end{figure}

\begin{figure}[!t]
\begin{center}
  \includegraphics[width=11cm]{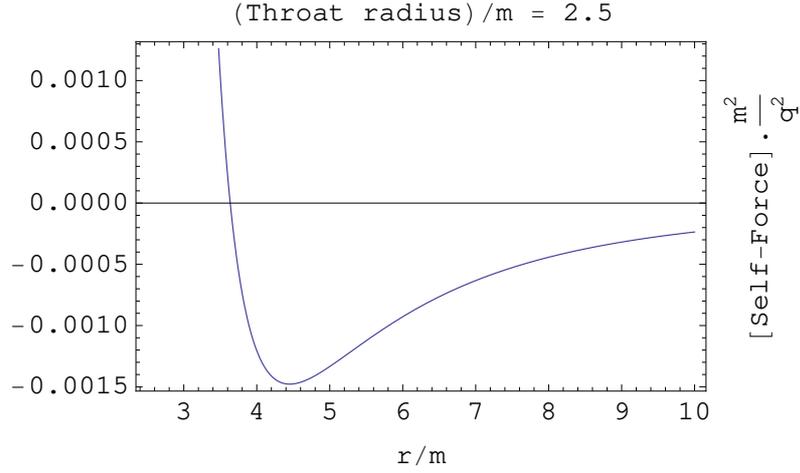}
  \caption{Dimensionless radial self-force ($\left(\mathrm{f}^{self}\right)\frac{m^2}{q^2}$) as a function of the dimensionless coordinate $r/m$ of a charge $q$ when the throat radius is at $c=2.5m$. The graph represent values for the position $r$ in the range $(c=2.5m,10m)$. Both regimes, repulsive and attractive, of the electrostatic self-force are observed in this case. }\label{2.5}
\end{center}
\end{figure}

\begin{figure}[!t]
\begin{center}
  \includegraphics[width=11cm]{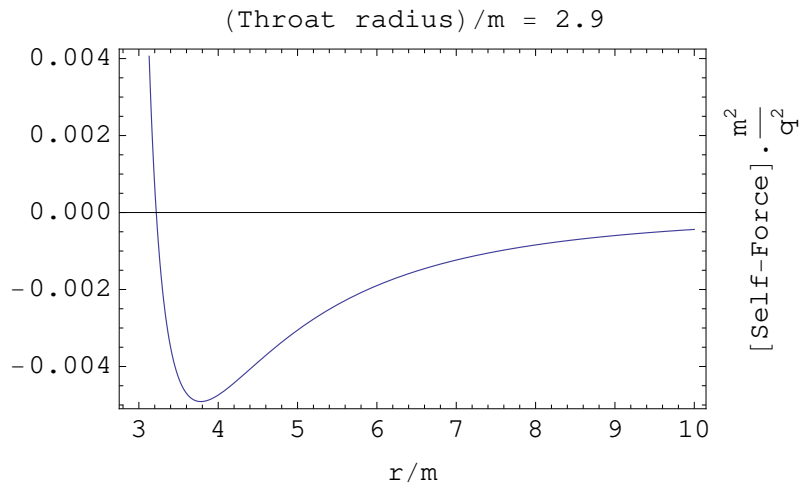}
  \caption{Dimensionless radial self-force ($\left(\mathrm{f}^{self}\right)\frac{m^2}{q^2}$) as a function of the dimensionless coordinate $r/m$ of a charge $q$ when the throat radius is at $c=2.9m$. The graph represent values for the position $r$ in the range $(c=2.9m,10m)$. This throat value is very close to the transition one, $c=3m$, beyond which only attractive regime is observe.}\label{2.9}
\end{center}
\end{figure}

To calculate the self-force we must consider the regular potential,
\be
V^{wh}_{ren}= \frac{qm}{ar} - q \sum_{l=0}^{\infty}  \left[\frac{(2l+1)}{2c^2f'_{l}(c)}+ g_{l}(c) \right]
\frac{f_{l}(a)}{f_{l}(c)}f_{l}(r) P_{l}(\cos\theta)\,.
\ee
The desired electrostatic self-force can be computed as that observed by a static observer at the position of the charge. This corresponds to the contravariant tetrad component of the force calculated with the renormalized potential $V^{wh}_{ren}$. Taking into account the definition of the electrostatic field (\ref{field}), we have
$$
\left(\mathrm{f}^{self}\right)^{(r)}=-qF^{(r)}_{\:\;\;(t)}u^{(t)}=-q ( \mathbf{e}^{(r)} )_r F^{r}_{\:\;\;t} u^t
= -q\sqrt{g_{rr}} g^{rr} F_{rt} \frac{1}{\sqrt{-g_{tt}}}
$$
\be\lb{fself}
=-q \partial_r V^{wh}_{ren}\,,
\ee
where $\mathbf{e}^{(r)}$ is the standard radial tetrad one-form, and the final result must be evaluated at the charge's position. One usually defines the electrostatic self-energy of the charge through the standard procedure
\be\lb{Uself}
U^{self}=\frac{q}{2}V^{wh}_{ren}(a)\,,
\ee
in order to obtain the force by
\be
\left(\mathrm{f}^{self}\right)^{(r)}=-\partial_a U^{self}\, ,
\ee
which justifies the definition (\ref{Uself}) for the self-energy. In the wormhole case, the electrostatic self-energy at a general position of the charge with radial coordinate $r$ is
\be\lb{U}
U^{self}=\frac{e^2m}{2r^2}-\frac{e^2}{2} \sum_{l=0}^{\infty}  \left[\frac{(2l+1)}{2c^2f'_{l}(c)}+ g_{l}(c) \right] \frac{f_{l}(r)^2}{f_{l}(c)}  \, ,
\ee
and the resulting electrostatic self-force is
\be\lb{force}
\left(\mathrm{f}^{self}\right)^{(r)}=\frac{q^2m}{r^3}+q^2\sum_{l}f'_{l}(r)\frac{f_{l}(r)}{f_{l}(c)}\left[\frac{2l+1}{2c^2f'_{l}(c)}+
g_{l}(c)\right] \,.
\ee
\begin{figure}[!t]
\begin{center}
  \includegraphics[width=11cm]{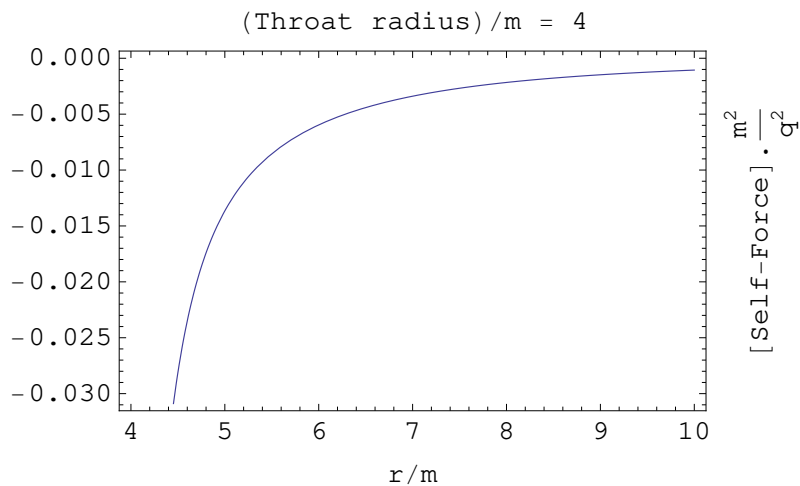}
  \caption{Dimensionless radial self-force ($\left(\mathrm{f}^{self}\right)\frac{m^2}{q^2}$) as a function of the dimensionless coordinate $r/m$ of a charge $q$ when the throat radius is at $c=4m$. The graph represent values for the position $r$ in the range $(c=4m,10m)$. Only attractive electrostatic self-force is possible in this case.}\label{4}
\end{center}
\end{figure}
The first term $q^2m/r^3$ is the self-force of the charge in Schwarzschild's black hole, while the remaining part is the correction due to the wormhole non trivial topology. The renormalization procedure had dealt with the divergent part which appeared only in the black hole term of the potential (\ref{ba2}), resulting in the known self-force $q^2m/r^3$ which was derived first in \cite{will80} and considered by several other authors for the Schwarzschild black hole (for instance in \cite{kpl} and \cite{vega}). The result (\ref{force}) for the thin-shell Schwarzschild wormhole is illustrated in the following Figures. The numerical analysis shows that the electrostatic self-force is always attractive towards the throat if the parameter $c$ of the wormhole throat radius is greater than $3m$ (Figures \ref{4} and \ref{3.1}). When $2m<c<3m$, the self-force may become repulsive if the charge is sufficiently proximate to the throat. This last observation is reflected in Figures \ref{2.5} and \ref{2.9}.

An important test for the correctness of the result is to take the limit $m\to 0$. In this situation, the wormhole space-time is flat everywhere except at the throat at $r=c$. This is the case of an infinitely short throat. Taking into account the limiting behavior of the radial independent functions;
\be
g_l(r) \xrightarrow{m \to 0}  r^l \quad and\;\quad f_l(r)\xrightarrow{m \to 0} 1/r^{l+1}\:,
\ee
the self-force (\ref{force}) for the flat wormhole with infinitely short throat is obtained:
\be
\mathrm{f}^{self}\xrightarrow{m \to 0} q^2\sum_{l=0}^{\infty}\frac{-(l+1)}{r^{l+2}}\frac{c^{l+1}}{r^{l+1}}\left[\frac{(2l+1)c^{l+2}}{-2(l+1)c^2}+
c^l\right]=\frac{q^2c}{r^3}\sum_{l=0}^{\infty}-(l+1)\frac{c^{l}}{r^{2l}}\left[\frac{-c^l}{-2(l+1)}\right]
\ee
$$
=
-\frac{q^2c}{2r^3}\sum_{l=0}^{\infty}\left[\left(\frac{c}{r}\right)^2\right]^l=
-\frac{q^2c}{2r^3}\left[\frac{1}{1-\left(c/r\right)^2}\right].
$$
The self-force is attractive towards the throat everywhere in the flat space-time, and gets infinitely large in the neighborhood of the throat where the curvature diverges. This result coincides with that of \cite{kb} and gives us confidence about the correctness of our solution. Similarly, if $r,c >> 2m$, the leading term is
\be
\mathrm{f}^{self} \sim -\frac{q^2c}{2r^3}\left[\frac{1}{1-\left(c/r\right)^2}+ \mathcal{O}\left( \frac{m}{c} + \frac{m}{r}\right)\right].
\ee
On the other hand, the asymptotic behavior ($r \to \infty$) is given by
\be\lb{asy}
\mathrm{f}^{self} \sim -\frac{q^2c}{2r^3}\left(1-\frac{2m}{c}\right)+\mathcal{O}\left(r^{-5}\right).
\ee
This expansion shows that the difference between the black hole and wormhole self-forces is manifested at leading-order, which is in contrast with other computations where the Schwarzschild black hole interior is replaced by a material body \cite{gralla}. Observe that in (\ref{asy}) the leading-order term vanishes as $c\to2m$ and the same happens at every order when the throat approaches the Schwarzschild radius. One might be tempted to say that there is no self-force induced for this value of throat radius, but in fact there is no consistent electrostatic field solution for one point charged particle when the throat is at $c=2m$. The results obtained above are only valid for throats which do not extend to the event horizon\footnote{Note that at $r=2m$ the independent radial functions $g_l(r)$ and $f_l^{'}(r)$ vanish and diverge respectively (for $l\neq0$), that's why many of the identities used to obtain the explicit potential expansion would not hold in the case $c=2m$.}. The reason is that the only electrostatic field on a Schwarzschild background which is well-behaved for $2m\leq r < \infty$ is spherically symmetric \cite{israel}. This means that the electro-vacuum part of the wormhole (i.e., region 2 in our notation, free of charge) can never adjust its potential, $V^{wh}_2$, to the boundary conditions at the throat imposed by the presence of the charge in the other region. The charged particle generates an equipotential surface at $c=2m$ which seems suitable, but the angular distribution of the electric field at the throat cannot be fulfilled by the spherically symmetric solution in the vacuum region. The only possibility to have consistent solutions in the case $c=2m$ would be the appearance of another charge in the free region, or work only with spherically symmetric distributions.

\section{Summary}
\setcounter{equation}{0}  % reset counter

A spherically symmetric thin-shell wormhole connecting two identical exterior Schwarzschild geometries is locally indistinguishable from a Schwarzschild black hole geometry. However, because the wormhole space-time presents a throat, there is an essential topological difference between both space-times which makes them globally very different. Following the proposal of our preceding work \cite{eoc12}, i.e. that the electrostatic self-force on a point charge could be used to probe the global aspects of a geometry, we have evaluated this force in both the black hole and the wormhole backgrounds. It was already known that the force on a charge in the vicinity of a Schwarzschild black hole pushes it away, for any position of the charge. In the case of a charge near the wormhole throat we have obtained an analytical expression of the self-force in the form of a series, which we have evaluated numerically. The results show that for a certain range of the parameters of the system the self-force is always attractive, that is, it points towards the wormhole throat for any position of the charge, while for another range of the parameters it can be attractive or repulsive depending on the position of the charge. The repulsive force only appears for a throat radius below $3m$, and for a charge placed very near from the wormhole throat. As the charge is placed far away, for any throat radius, the self-force is always attractive. This can be observed directly from the asymptotic expansion (\ref{asy}). Thus the electrostatics of a point charge would allow to decide whether the background geometry presents a throat or not. A detail to be noted is that now, in the case of a charge in a spherical wormhole geometry, the self-force on a charge placed far from the throat is always attractive, while in the case studied in \cite{eoc12} the self-force on a charge very far from the throat of a cylindrical wormhole with a deficit angle, was always repulsive. This can be understood as a consequence of the different asymptotic behaviors of the Schwarzschild and cosmic string geometries. While the Schwarzschild space-time is asymptotically flat (as $r\to \infty$ the metric becomes Minkowski), the angle deficit associated to a gauge cosmic string is constant. Then, very far from the throat of a cylindrical wormhole, the dominant effect on the electric field lines is that of the deficit angle which induces a repulsive force. In the spherically symmetric wormhole connecting two Schwarzschild geometries, instead, this effect is not present as the metric tends to that of a flat background. Near the throat, for the cylindrical wormhole associated to a cosmic string, the attractive effect of the throat reverses the repulsive deficit angle effect deriving in a force towards the hole. For the thin-shell Schwarzschild wormhole, the intense repulsive effect in the exterior neighboring geometry of a black hole horizon may reverse the attractive contribution of the wormhole.

\section*{Appendix}  % use *-form to suppress numbering
\renewcommand{\theequation}{A.\arabic{equation}}% redefine the command that creates the equation no.

\setcounter{equation}{0}  % reset counter

In this appendix it will be shown that the expansion (\ref{Vcw}) is equivalent to the electrostatic potential (\ref{willy}) for a Schwarzschild black hole. The arguments are analogous to those in \cite{copson}, the main difference is that the electrostatic potential of that reference doesn't fit the correct boundary conditions at infinity. For this purpose it may be convenient to introduce the so called isotropic radial coordinate $\rho$ given in terms of the Schwarzschild one $r$ by
the following formula
\be\lb{ro}
\rho=\frac{r-m+\sqrt{r(r-2m)}}{2},\qquad r>2m.
\ee
Note that this coordinate do not cover completely the black hole space-time, only the outer region of the horizon at $r=2m$.
The previous expression may be inverted to give
\be\lb{ro2}
r=\rho\bigg(1+\frac{m}{2\rho}\bigg)^2,\qquad 2\rho>m.
\ee
The Schwarzschild distance element in isotropic coordinates takes the following form
\be
g_4=-\frac{[1-(m/2\rho)]^2}{[1+(m/2\rho)]^2}dt^2+\bigg(1+\frac{m}{2\rho}\bigg)^4(d\rho^2+\rho^2d\theta^2+\rho^2\sin^2\theta d\varphi^2).
\ee
The electrostatic potential for a point charge in the black hole, (\ref{willy}), is expressed in isotropic coordinates as
$$
V^{bh}=\frac{q}{[1+(m/2b)]^2\rho[1+(m/2\rho)]^2}\bigg[\sqrt{\frac{\rho^2+b_1^2-2b_1\rho \cos\theta}{\rho^2+b^2-2b\rho \cos\theta}}+\frac{m^2}{4b^2}\sqrt{\frac{\rho^2+b^2-2b\rho \cos\theta}{\rho^2+b_1^2-2b_1\rho \cos\theta}}\bigg]
$$
\be
+\frac{q m}{b}\frac{1}{[1+(m/2b)]^2}\frac{1}{\rho[1+(m/2\rho)]^2},
\ee
with $b=\frac{a-m+\sqrt{a(a-2m)}}{2}$ and $b_1=\frac{m^2}{4b}$. The task is to show that the expansion of this potential in the independent functions (\ref{fyg}) is exactly (\ref{Vcw}). By further defining the dimensionless coordinate $\eta=2\rho/m$ and the dimensionless position of the charge as $\beta=2b/m$
the potential reads as follows
\be
V^{bh}=\frac{4q\eta\beta}{2m[1+\eta]^2[1+\beta]^2}\bigg[\sqrt{\frac{\eta^2+\beta^2-2\beta\eta \cos\theta}{\beta^2\eta^2+1-2\beta\eta \cos\theta}}+\sqrt{\frac{\beta^2\eta^2+1-2\beta\eta \cos\theta}{\eta^2+\beta^2-2\beta\eta \cos\theta}}\bigg]
+\frac{4 q \beta \eta}{m[1+\eta]^2[1+\beta]^2}.
\ee
At $\theta=0$ and $r>a$ this expression reduces to
\be\lb{poto}
V^{bh}=V^{bh}_1+V^{bh}_2=\frac{4q\eta\beta}{2m[1+\eta]^2[1+\beta]^2}\bigg[\frac{\eta-\beta}{\beta\eta-1}+\frac{\beta\eta-1}{\eta-\beta}\bigg]
+\frac{4 q \beta \eta}{m[1+\eta]^2[1+\beta]^2}.
\ee
The first term of (\ref{poto}) can be worked out in terms of the geometric series as follows
$$
V^{bh}_1=\frac{2q\beta}{m(\beta+1)^2}\frac{\eta-1}{\eta+1}\frac{\eta}{1-\eta^2}\bigg[\frac{1}{\beta}\bigg(1-\frac{\beta}{\eta}\bigg)\bigg(1-\frac{1}
{\beta\eta}\bigg)^{-1}+\frac{1}{\beta}\bigg(1-\frac{1}{\beta\eta}\bigg)\bigg(1-\frac{\beta}{\eta}\bigg)^{-1}\bigg]
$$
\be\lb{poto2}
=\frac{2q\beta}{m(\beta+1)^2}\frac{\eta-1}{\eta+1}\frac{1}{\eta}\bigg(1+\frac{1}{\eta^2}+\frac{1}{\eta^4}+..\bigg)\bigg[\bigg(\beta+\frac{1}{\beta}\bigg)+\bigg(\beta^2-2+\frac{1}{\beta^2}\bigg)\frac{1}{\eta}+..]
\ee
$$
=\frac{2q\beta}{m(\beta+1)^2}\frac{\eta-1}{\eta+1}\bigg[\bigg(\beta+\frac{1}{\beta}\bigg)\frac{1}{\eta}+\bigg(\beta^2-2+\frac{1}{\beta^2}\bigg)\frac{1}{\eta^2}+\bigg(\beta^3+\frac{1}{\beta^3}\bigg)\frac{1}{\eta^3}+
\bigg(\beta^4-2+\frac{1}{\beta^4}\bigg)\frac{1}{\eta^4}+..].
$$
Now, in order to connect this with the Legendre functions $P_{l}$ and $Q_l$ the three following identities will be useful
\be\lb{ww1}
\frac{1}{(x+\sqrt{x^2-1})^n}=-n\sum_{m=0}^{\infty}\frac{(4m+2n-1)}{4\pi}\frac{\Gamma(m-1/2)\Gamma(m+n-1/2)}{m! (m+n)!}Q_{2m+n-1}(x),
\ee
$$
(\beta^2+2-\frac{1}{\beta^2})P'_{2n}(\frac{1}{2}(\beta+\frac{1}{\beta}))=-\frac{2n(2n+1)}{\pi}\sum_{k=0}^{\infty}\bigg(\beta^{2n-2k+1}+\frac{1}{\beta^{2n-2k+1}}\bigg)
$$
\be\lb{ww2}
\times \frac{\Gamma(k-\frac{1}{2})\Gamma(2n-k+\frac{1}{2})}{k!(2n-k+1)!},
\ee
$$
(\beta^2+2-\frac{1}{\beta^2})P'_{2n+1}(\frac{1}{2}(\beta+\frac{1}{\beta}))=-\frac{(2n+1)(2n+2)}{\pi}\sum_{k=0}^{\infty}\bigg(\beta^{2n-2k+2}-2+\frac{1}{\beta^{2n-2k+2}}\bigg)
$$
\be\lb{ww3}
\times \frac{\Gamma(k-\frac{1}{2})\Gamma(2n-k+\frac{3}{2})}{k!(2n-k+2)!}.
\ee
In addition, by taking the derivative with respect to $x$ of (\ref{ww1}), it follows that
\be\lb{ww4}
\frac{1}{(\sqrt{x^2-1})(x+\sqrt{x^2-1})^n}=-n\sum_{m=0}^{\infty}\frac{(4m+2n-1)}{4\pi}\frac{\Gamma(m-1/2)\Gamma(m+n-1/2)}{m! (m+n)!}Q'_{2m+n-1}(x).
\ee
These identities can be implemented in (\ref{poto2}) by evaluating
\be
x=\frac{\eta^2+1}{2\eta},
\ee
in (\ref{ww4}). By comparing with (\ref{poto2}) it follows that
$$
V^{bh}_1=\frac{q\beta(\eta-1)^2}{2m\eta(\beta+1)^2}\bigg[\sum_{m=0}^{\infty}\sum_{p=0}^{\infty}\frac{(4m+4p+1)}{4\pi}\bigg(\beta^{2m+1}+\frac{1}{\beta^{2m+1}}\bigg)\frac{\Gamma(p-\frac{1}{2})\Gamma(2m+p+\frac{1}{2})}{p!(2m+p+1)!}
$$
$$
\times Q'_{2m+2p}(\frac{\eta^2+1}{2\eta})
+\sum_{m=0}^{\infty}\sum_{p=0}^{\infty}\frac{(4m+4p+3)}{4\pi}\bigg(\beta^{2m+2}-2+\frac{1}{\beta^{2m+2}}\bigg)\frac{\Gamma(p-\frac{1}{2})\Gamma(2m+p+\frac{3}{2})}{p!(2m+p+1)!}
$$
$$
\times Q'_{2m+2p+1}(\frac{\eta^2+1}{2\eta})\bigg]
=\frac{e\beta}{m(\beta+1)^2}\frac{(\eta-1)^2}{2\eta}\bigg[\sum_{n=0}^{\infty}\sum_{t=0}^{n}\frac{(4n+3)}{4\pi}\bigg(\beta^{2(n-t)+1}+\frac{1}{\beta^{2(n-t)+1}}\bigg)
$$
$$
\times \frac{\Gamma(t-\frac{1}{2})\Gamma(2n-t+\frac{3}{2})}{t!(2n-t+2)!}Q'_{2n+1}(\frac{\eta^2+1}{2\eta})
+\sum_{n=0}^{\infty}\sum_{t=0}^{n}\frac{(4n+1)}{4\pi}\bigg(\beta^{2(n-t+1)}-2+\frac{1}{\beta^{2(n-t+1)}}\bigg)
$$
\be
\times\frac{\Gamma(t-\frac{1}{2})\Gamma(2n-t+\frac{1}{2})}{t!(2n-t+1)!}Q'_{2n}(\frac{\eta^2+1}{2\eta})-\frac{1}{2}\bigg(\beta+\frac{1}{\beta}\bigg)Q_0'(\frac{\eta^2+1}{2\eta})\bigg].
\ee
The last expression can be worked out further by use of (\ref{ww2})-(\ref{ww3}) to give
$$
V^{bh}_1=\frac{q\beta}{m(\beta+1)^2}\frac{(\eta-1)^2}{4\eta}\bigg[\bigg(\beta+\frac{1}{\beta}\bigg)Q_0'(\frac{\eta^2+1}{2\eta})+\frac{1}{2}\sum_{n=1}^{\infty}\frac{2n+1}{n(n+1)}\bigg(\beta^2-2+\frac{1}{\beta^2}\bigg)
$$
\be\lb{lost}
\times P'_{n}(\beta)Q'_n(\frac{\eta^2+1}{2\eta})\bigg].
\ee
By expressing (\ref{lost}) in terms the Schwarzschild coordinates $r$ and $a$ it follows that
$$
V^{bh}_1=\frac{q(r-2m)}{am}Q'_0\bigg(\frac{r}{m}-1\bigg)-q\bigg[\frac{(r-2m)}{m^2}Q'_0\bigg(\frac{r}{m}-1\bigg)+\frac{1}{m^3}\sum_{n=1}^{\infty}\frac{2n+1}{n(n+1)}
$$
\be
\times(a-2m)P'_{n}\bigg(\frac{a}{m}-1\bigg)(r-2m)Q'_n\bigg(\frac{r}{m}-1\bigg)\bigg].
\ee
As we pointed earlier, the Legendre functions which appear in this section are been differentiated in its argument and then evaluated in the specified value, so that the elementary identity
\be
Q'_0\bigg(\frac{r}{m}-1\bigg)=\frac{d Q_0}{dx}\mid_{(\frac{r}{m}-1)}=-\frac{m^2}{r(r-2m)},
\ee
together with the definitions (\ref{fyg}), shows that:
\be
V^{bh}_1=-\frac{qm}{ar}+q\sum_{l=0}^{\infty} g_{l}(a)f_{l}(r).
\ee
On the other hand the expression of the potential $V^{bh}_2$ in (\ref{poto2}) in Schwarzschild coordinates is
\be
V^{bh}_2=\frac{qm}{ar},
\ee
and therefore
\be\lb{redu}
V^{bh}=V^{bh}_1+V^{bh}_2=q\sum_{l=0}^{\infty} g_{l}(a)f_{l}(r).
\ee
This is the electrostatic potential evaluated at the surface $\theta=0$. Since we know that for $\theta\neq 0$ the expression should be an expansion of the form
\be
V(r,\theta)=q\sum_{l=0}^{\infty} F_l(r)P_l(\cos\theta),
\ee
with $F_l(r)$ radial functions labelled by $l$ and which are solutions of (\ref{lapli}). This expression should reduce to  (\ref{redu}) when $\theta=0$, or equivalently for $P_l(1)=1$. From this it follows that the full expression for the potential should be
\be
V^{bh}=V^{bh}_1+V^{bh}_2=q\sum_{l=0}^{\infty} g_{l}(a)f_{l}(r)P_l(\cos\theta),
\ee
which is the formula (\ref{Vcw}) described in the text for $r>a$. The corresponding expression for $r<a$
is completely analogous and we just omit it. Therefore we have seen that the expansion (\ref{Vcw}) is the same as the potential (\ref{willy}), which is what we wanted to show.
\\

{\bf Acknowledgements:}  The authors are supported by CONICET (Argentina).
\\

\end{document}